\documentclass[%
 reprint,
superscriptaddress,
 amsmath,amssymb,
 aps,
pra,
]{revtex4-2}

\usepackage{graphicx}
\usepackage{dcolumn}
\usepackage{bm}
\usepackage{hyperref}
\usepackage{xcolor}
\usepackage{soul}
\usepackage{placeins}

\begin{document}
\preprint{APS/123-QED}


\title{Benchmarking Many-body Approaches for the Determination of Isotope Shift Constants: Application to the Li, Be$^+$ and Ar$^{15+}$ Isoelectronic Systems}

\author{B. K. Sahoo}
\email{bijaya@prl.res.in}
\affiliation{Atomic and Molecular Physics Division, Physical Research Laboratory, Navrangpura, Ahmedabad 380009, India}
\author{B. Ohayon}
\email{bohayon@ethz.ch}
\affiliation{
Institute for Particle Physics and Astrophysics, ETH Z\"urich, CH-8093 Z\"urich, Switzerland 
}

\date{\today}

\begin{abstract}
We have applied relativistic coupled-cluster (RCC) theory to determine the isotope shift (IS) constants of the first eight low-lying states of the Li, Be$^+$ and Ar$^{15+}$ isoelectronic systems. Though the RCC theory with singles, doubles and triples approximation (RCCSDT method) is an exact method for these systems for a given set of basis functions, we notice large differences in the results from this method when various procedures in the RCC theory framework are adopted to estimate the IS constants. This has been demonstrated by presenting the IS constants of the aforementioned states from the finite-field, expectation value and analytical response (AR) approaches of the RCCSDT method. Contributions from valence triple excitations, Breit interaction and lower-order QED effects to the evaluation of these IS constants are also highlighted. Our results are compared with high-precision calculations reported using few-body methods wherever possible. We find that results from the AR procedure are more reliable than the other two approaches. This analysis is crucial for understanding the roles of electron correlation effects in the accurate determination of IS constants in the heavier atomic systems, where few-body methods cannot be applied.
\end{abstract}

\maketitle

\section{Introduction}

There have been a lot of demand to develop many-body methods for the evaluation of isotope shift (IS) constants in atomic systems, which in combination with measurements offer information about the nuclear charge radii of atomic nuclei \cite{Otten,2012-Cheal}. These model-independent nuclear charge radii are used to validate different nuclear methods \cite{Kosz, 2008-ISO, 2018-RadialOverlap}, and shed light on the nuclear many-body problem \cite{Campbell,Garcia, Miller,Kosz}. 

Owing to tiny magnitudes, estimating ISs at the first-order perturbation is good enough for comparing with most of the measurements. At this approximation, contribution to the IS is divided into two parts: field-shift (FS) and mass-shift (MS). The MS is further divided into normal mass shift (NMS) and specific mass shift (SMS). Theoretically, these shifts are estimated by determining their respective constants. The FS and NMS constants are described by one-body operators, while the SMS constant is described by a two-body operator. Thus, accurate calculation of the SMS constants is generally presumed to be more challenging than evaluating the FS and NMS constants. However, the NMS constants are usually estimated through a scaling procedure \cite{Otten,Kosz, 2018-RadialOverlap,Campbell,Garcia, Miller} by adopting Virial theorem \cite{Fock}, which relates kinetic energy with total energy of an atomic system. In some cases, the potential of the many-body methods employed for the determination of the SMS constants and FS constants could be tested by comparing calculations with the corresponding measurements; where the measured values are inferred with the help of a King plot \cite{Kings}. 

There have been attempts to describe ISs using many-body calculations, but significant differences are observed among the experimental values and the results from the many-body calculations 
 \cite{Martensson,Safronova,Berengut,Korol,Sahoo2010,Naze,Roy,Yerokhin}. In fact, often high-accuracy calculations are claimed for the FS constants to infer the nuclear charge radii but the same methods show large deviations of the NMS constants from the their scaled values. This may seriously question the validity of the inferred nuclear charge radii, as well as the potential of the employed methods. In this view, it is imperative to systematically analyze the approaches and the many-body methods considered to estimate the IS constants.

Generally, the calculated energies from a particular method compare well with the corresponding experimental values, validating the accuracy of the atomic wave functions. However, this condition is not sufficient to ascertain the accuracy of the calculated IS constants, as is shown in this work. Customarily, wave functions of a multi-electron system are obtained at least in two steps; First a mean-field approach is adopted and then, the neglected electron correlation effects are included systematically. Configuration interaction (CI) method and its variants are typically employed to evaluate the IS constants of atomic systems \cite{Naze,Yerokhin}. However, it is not possible to employ a full CI method to an atomic system with more than three electrons using the available computational facilities even with a reasonable size of basis function. Coupled-cluster (CC) theory is considered to be a more potent many-body method compared to a CI method at a given level of truncation for calculating wave functions of multi-electron systems. This is because it accounts for size-extensitivity and size-consistent behavior, and captures more physical effects due to the electron correlations \cite{Bishop,Bartlett} than the truncated CI methods at similar level of approximation.

There are many variants of CC methods proposed in the literature for accurate determination of atomic properties (e.g. see Refs. \cite{Bishop,Monkhorst}). However, these methods are not explored extensively while reporting the IS constants to examine the reliability of the calculations. The relativistic version of CC (RCC) theory is a natural choice to apply for accurate determination of the IS constants. Furthermore, in most cases many-body methods are applied in the finite-field (FF) framework for the estimations of the IS constants \cite{Berengut,Korol,Naze,Roy,Yerokhin}. In this work, we analyze the IS constants (i.e. the FS, NMS and SMS constants) by performing many-body calculations in three different procedures; namely the aforementioned FF, a direct expectation value evaluation (EVE), and analytical response (AR) approaches. We have thus adopted the RCC theory framework to formulate these three approaches.

The many-body theories like CI and RCC methods are apt for determining properties of atomic systems having more than three electrons, while one can find other methods that are more suitable for high-precision calculations in a few electron atomic systems \cite{Drake,Pachucki,Yan,Wang}. It is, however, convenient to apply a newly developed many-body approach to the lithium (Li) atom and its isoelectronic atomic systems to benchmark potential of the method to incorporate electron correlation and relativistic effects \cite{Li,Yu}. This is owing to the fact that a many-body theory becomes an exact method with singles, doubles and triples approximation with a given set of basis functions. Besides, a number of analysis can be performed on these systems within a stipulated time frame due to their small sizes. For this purpose, we have considered here the neutral Li atom, singly charged beryllium (Be$^+$) and highly charged Li-like argon (Ar$^{15+}$) ion to analyze their IS shifts. Also, we calculate the IS constants for as many as eight low-lying states of the above systems so that it can offer more insight into the role of electron correlation effects in the atomic systems by analysing trends in the results of the states belonging to different angular momentum and level positioning.

\begin{table}[t]
\caption{Calculated energies (in cm$^{-1}$) of the first eight low-lying states of the Li atom with infinite nuclear mass, using the DC, DCB and DCQ Hamiltonians. Results are given from the DHF, RMP(2), RCCSD and full RCC methods to demonstrate the roles of electron correlation effects in the evaluation of energies of the above states.}
\begin{ruledtabular}
\begin{tabular}{lrrrr}
State  & DHF  & RMP(2)  & RCCSD & RCC \\
\hline \\
\multicolumn{5}{c}{DC Hamiltonian} \\ 
$2s ~ ^2S_{1/2}$  & 43087.33 & 43444.25 & 43483.17 & 43488.23 \\
$2p ~ ^2P_{1/2}$  & 28232.86 & 28530.50 & 28577.89 & 28581.89 \\
$2p ~ ^2P_{3/2}$  & 28232.30 & 28529.85 & 28577.22 & 28581.19 \\
$3s ~ ^2S_{1/2}$  & 16197.33 & 16272.86 & 16280.59 & 16281.65 \\
$3p ~ ^2P_{1/2}$  & 12459.93 & 12547.23 & 12560.13 & 12561.34 \\
$3p ~ ^2P_{3/2}$  & 12459.76 & 12547.02 & 12559.92 & 12561.13 \\
$3d ~ ^2D_{3/2}$  & 12194.37 & 12203.26 & 12204.86 & 12204.94 \\
$3d ~ ^2D_{5/2}$  & 12194.33 & 12203.22 & 12204.82 & 12204.90 \\
& & & & \\
\multicolumn{5}{c}{DCB Hamiltonian} \\ 
$2s ~ ^2S_{1/2}$  & 43086.30 & 43443.54 & 43482.39 & 43487.42 \\
$2p ~ ^2P_{1/2}$  & 28232.39 & 28530.13 & 28577.49 & 28581.47 \\
$2p ~ ^2P_{3/2}$  & 28232.14 & 28529.73 & 28577.06 & 28581.01\\
$3s ~ ^2S_{1/2}$  & 16197.09 & 16272.70 & 16280.41 & 16281.51 \\
$3p ~ ^2P_{1/2}$  & 12459.77 & 12547.10 & 12559.99 & 12561.19 \\
$3p ~ ^2P_{3/2}$  & 12459.70 & 12546.98 & 12559.87 & 12561.07 \\
$3d ~ ^2D_{3/2}$  & 12194.37 & 12203.26 & 12204.86 & 12204.94 \\
$3d ~ ^2D_{5/2}$  & 12194.33 & 12203.22 & 12204.83 & 12204.90 \\
& & & & \\
\multicolumn{5}{c}{DCQ Hamiltonian} \\ 
$2s ~ ^2S_{1/2}$  & 43087.00 & 43443.93 & 43482.86 & 43487.71 \\
$2p ~ ^2P_{1/2}$  & 28232.91 & 28530.56 & 28577.96 & 28581.95 \\
$2p ~ ^2P_{3/2}$  & 28232.35 & 28529.90 & 28577.28 & 28581.25 \\
$3s ~ ^2S_{1/2}$  & 16197.25 & 16272.79 & 16280.52 & 16281.58 \\
$3p ~ ^2P_{1/2}$  & 12459.95 & 12547.25 & 12560.15 & 12561.36 \\
$3p ~ ^2P_{3/2}$  & 12459.77 & 12547.04 & 12559.94 & 12561.15 \\
$3d ~ ^2D_{3/2}$  & 12194.36 & 12203.26 & 12204.86 & 12204.94 \\
$3d ~ ^2D_{5/2}$  & 12194.33 & 12203.22 & 12204.82 & 12204.90 \\
\end{tabular}
\end{ruledtabular}
\label{tabLi}
\end{table}

\begin{table}[t]
\caption{Calculated energies (in cm$^{-1}$) of the low-lying states of the Be$^+$ ion with infinite nuclear mass. Results are given from the DHF, RMP(2), RCCSD and full RCC methods using the DC, DCB and DCQ Hamiltonians.}
\begin{ruledtabular}
\begin{tabular}{lrrrr}
State  & DHF  & RMP(2)  & RCCSD & RCC \\
\hline \\
\multicolumn{5}{c}{DC Hamiltonian} \\ 
$2s ~ ^2S_{1/2}$  &  146210.22  &  146836.83  & 146884.65 & 146889.91 \\
$2p ~ ^2P_{1/2}$  &  114005.30  &  114856.96  & 114943.66 & 114948.36 \\
$2p ~ ^2P_{3/2}$  &  113996.39  &  114847.32  & 114933.93 &  114938.42 \\
$3s ~ ^2S_{1/2}$  &  58494.75   &   58641.04  &  58652.34 &   58653.91 \\
$3p ~ ^2P_{1/2}$  &  50133.32   &   50363.54  &  50385.07 &   50386.46 \\
$3p ~ ^2P_{3/2}$  &  50130.64   &   50360.65  &  50382.17 &   50383.53 \\
$3d ~ ^2D_{3/2}$  &  48788.33   &   48826.04  &  48830.74 &   48830.76 \\
$3d ~ ^2D_{5/2}$  &  48787.75   &   48825.46  &  48830.15 &   48830.17 \\
& & & & \\
\multicolumn{5}{c}{DCB Hamiltonian} \\ 
$2s ~ ^2S_{1/2}$  &  146205.03  &  146832.81 & 146880.44  &  146885.59 \\
$2p ~ ^2P_{1/2}$  &  114000.23  &  114852.39 & 114938.94  &  114943.56 \\
$2p ~ ^2P_{3/2}$  &  113994.43  &  114845.52 & 114931.96  &  114936.36 \\
$3s ~ ^2S_{1/2}$  &  58493.44   &    58640.04 &  58651.30 &  58652.83 \\
$3p ~ ^2P_{1/2}$  &  50131.73   &    50362.09 &  50383.58 &  50384.94 \\
$3p ~ ^2P_{3/2}$  &  50130.03   &    50360.08 &  50381.55 &  50382.89 \\
$3d ~ ^2D_{3/2}$  &  48788.32   &    48826.08 &  48830.78 &  48830.80 \\
$3d ~ ^2D_{5/2}$  &  48787.75   &    48825.48 &  48830.18 &  48830.20 \\
& & & & \\
\multicolumn{5}{c}{DCQ Hamiltonian} \\ 
$2s ~ ^2S_{1/2}$  &  146208.71 &  146835.35  &  146883.16 &  146889.36 \\
$2p ~ ^2P_{1/2}$  &  114005.57 &  114857.28  &  114943.98 &  114948.68 \\
$2p ~ ^2P_{3/2}$  &  113996.63 &  114847.60  &  114934.23 &  114938.71 \\
$3s ~ ^2S_{1/2}$  &  58494.36  &  58640.66  &  58651.96 &  58653.81 \\
$3p ~ ^2P_{1/2}$  &  50133.40  &  50363.63  &  50385.17 &  50386.55 \\
$3p ~ ^2P_{3/2}$  &  50130.71  &  50360.73  &  50382.25 &  50383.61 \\
$3d ~ ^2D_{3/2}$  &  48788.33  &  48826.05  &  48830.74 &  48830.77 \\
$3d ~ ^2D_{5/2}$  &  48787.75  &  48825.46  &  48830.15 &  48830.18 \\
\end{tabular}
\end{ruledtabular}
\label{tabBe}
\end{table}

\begin{table}[t]
\caption{Calculated energies (in cm$^{-1}$) of the low-lying states of Ar$^{15+}$ with infinite nuclear mass, using the DHF, RMP(2), RCCSD and full RCC methods. Results from the DC, DCB and DCQ Hamiltonians are quoted separately for the comparison.}
\begin{ruledtabular}
\begin{tabular}{lrrrr}
State  & DHF  & RMP(2)  & RCCSD & RCC \\
\hline \\
\multicolumn{5}{c}{DC Hamiltonian} \\ 
$2s ~ ^2S_{1/2}$  &  7408273 & 7409478 & 7409503 & 7409503 \\
$2p ~ ^2P_{1/2}$  &  7150039 & 7152427 & 7152481 & 7152482 \\
$2p ~ ^2P_{3/2}$  &  7123169 & 7125492 & 7125546 & 7125546 \\
$3s ~ ^2S_{1/2}$  &  3231886 & 3231886 & 3232228 & 3232228 \\
$3p ~ ^2P_{1/2}$  &  3160559 & 3161141 & 3161154 & 3161154 \\
$3p ~ ^2P_{3/2}$  &  3152626 & 3153192 & 3153205 & 3153206 \\
$3d ~ ^2D_{3/2}$  &  3125974 & 3126175 & 3126181 & 3126180 \\
$3d ~ ^2D_{5/2}$  &  3123594 & 3123793 & 3123799 & 3123798 \\
& & & & \\
\multicolumn{5}{c}{DCB Hamiltonian} \\ 
$2s ~ ^2S_{1/2}$  &  7406909 &  7408150 & 7408174 & 7408174 \\
$2p ~ ^2P_{1/2}$  &  7147450 &  7149863 & 7149915 & 7149915 \\
$2p ~ ^2P_{3/2}$  &  7122015 &  7124339 & 7124391 & 7124391 \\
$3s ~ ^2S_{1/2}$  &  3231503 &  3231848 & 3231856 & 3231857 \\
$3p ~ ^2P_{1/2}$  &  3159830 &  3160417 & 3160430 & 3160430 \\
$3p ~ ^2P_{3/2}$  &  3152312 &  3152878 & 3152891 & 3152891 \\
$3d ~ ^2D_{3/2}$  &  3125950 &  3126157 & 3126163 & 3126163 \\
$3d ~ ^2D_{5/2}$  &  3123587 &  3123789 & 3123795 & 3123795 \\
& & & & \\
\multicolumn{5}{c}{DCQ Hamiltonian} \\ 
$2s ~ ^2S_{1/2}$  &  7407530 & 7408735 & 7408760 & 7408760 \\
$2p ~ ^2P_{1/2}$  &  7150039 & 7152427 & 7152482 & 7152482 \\
$2p ~ ^2P_{3/2}$  &  7123185 & 7125508 & 7125562 & 7125562 \\
$3s ~ ^2S_{1/2}$  &  3231671 & 3232005 & 3232014 & 3232014 \\
$3p ~ ^2P_{1/2}$  &  3160558 & 3161140 & 3161154 & 3161154 \\
$3p ~ ^2P_{3/2}$  &  3152630 & 3153196 & 3153210 & 3153210 \\
$3d ~ ^2D_{3/2}$  &  3125973 & 3126174 & 3126180 & 3126179 \\
$3d ~ ^2D_{5/2}$  &  3123594 & 3123792 & 3123799 & 3123798 \\
\end{tabular}
\end{ruledtabular}
\label{tabAr}
\end{table}

\section{Theory}

The first-order IS of an atomic transition $\delta \nu$ between elements $A$ and $A'$ can be determined by \cite{Seltzer,Ahmad}
\begin{equation}
\delta v^{A,A'} = F^{\text{FS}} \Lambda^{A,A'} + K^{\text{MS}}(\mu_A-\mu_{A'}), 
\label{IS}
\end{equation}
where $F^{\text{FS}}$ and $K^{\text{MS}}$ are known as the FS and MS constants, respectively, that can be determined by performing atomic calculations, and $\mu_A=\frac{1}{M_A+m_e}$ with mass of an electron $m_e$, and nuclear mass $M_A$. The MS constant is given by $K^{\text{MS}}=K^{\text{NMS}} +K^{\text{SMS}}$ where $K^{\text{NMS}}$ and $K^{\text{SMS}}$ are the NMS and SMS constants respectively. The nuclear factor $\Lambda^{A,A'}$ is given by
\begin{eqnarray}
 \Lambda^{A,A'} = \sum_k \frac{C_k}{C_1} \delta \langle r^{2k} \rangle^{A,A'},
 \end{eqnarray}
where $C$s are known as the Seltzer coefficients that can be evaluated by the nuclear calculation. In the present work, we have considered only $k=1$ to estimate $F^{\text{FS}}=\langle O^{\text{FS}} \rangle$ for a given atomic state by defining
\begin{eqnarray}
O^{\text{FS}} &=& - \sum_i \frac{\partial V_n(\langle r^2 \rangle,r_i)}{\partial \langle r^2 \rangle}, 
\end{eqnarray}
where $V_n$ is the nuclear potential seen by an electron and $\langle r^2 \rangle$ is the root mean square (rms) nuclear charge radius.

Similarly, we evaluate the NMS and SMS constants as $K^{\text{NMS}}= \langle O^{\text{NMS}} \rangle$ and $K^{\text{SMS}}=\langle O^{\text{SMS}} \rangle$, where in the relativistic theory, $O^{\text{NMS}}$ and $O^{\text{SMS}}$ take the form \cite{Palmer_1987,1997-Shab}
\begin{eqnarray}
O^{\text{NMS}} &=& \frac{1}{2}\sum_i \left ({\vec p}_i^{~2} - \frac{\alpha Z}{r_i} {\vec \alpha}_i^D \cdot {\vec p}_i \right. \nonumber \\ && \left. - \frac{\alpha Z}{r_i} ({\vec \alpha}_i^D \cdot {\vec C}_i^1){\vec C}_i^1 \cdot {\vec p}_i \right ),
\label{nmsexp}
\end{eqnarray}
and
\begin{eqnarray}
O^{\text{SMS}} &=& \frac{1}{2} \sum_{i\ne j} \left ({\vec p}_i \cdot {\vec p}_j - \frac{\alpha Z}{r_i} {\vec \alpha}_i^D \cdot {\vec p}_j \right. \nonumber \\ && \left. - \frac{\alpha Z}{r_i} ({\vec \alpha}_i^D \cdot {\vec C}_i^1) ({\vec p}_j \cdot {\vec C}_j^1) \right ),
\end{eqnarray}
respectively, where $\alpha$ is the fine-structure constant, $Z$ is the atomic number and $\alpha^D$ is the Dirac operator.

As can be seen from the expression given by Eq. (\ref{nmsexp}), the NMS operator expression contains $p^2$. The Virial theorem suggests kinetic energy is equal to negative of the total energy in a spherically symmetric system \cite{bransden}. In the scaling approach, therefore, the NMS constant is given by $K^{\text{NMS}} \simeq m_e E^{\text{expt}}$ with the experimental energy of the state $E^\text{{expt}}$. As seen from the above equation, the NMS operator has more terms and the relativistic form of the kinetic energy is slightly different than $p^2/2m_e$. Moreover, only $S$ states are spherically symmetric in atomic systems, and so the NMS constants of these states may obey the Virial theorem. There could be deviation from this scaling in the other states, which needs to be probed by performing many-body calculations of the NMS constants. 

\begin{table*}[t]
\caption{Comparison of the final values of the calculated electron affinity energies (in cm$^{-1}$) with experimental values for the first eight low-lying states of the $^7$Li, $^9$Be$^+$ and $^{40}$Ar$^{15+}$. The ground state electron affinity of Ar$^{15+}$, which appears in italics, is taken from another theoretical calculation (see text). $\delta$ represents the fractional difference (in \%) between our calculations and experimental data, with experimental uncertainty given in parenthesis when it is appreciable. The results for Ar$^{15+}$ should be read as $\times 10^{3}$.}
\begin{ruledtabular}
\begin{tabular}{l rrr rrr rrr}
  & \multicolumn{3}{c}{$^7$Li} & \multicolumn{3}{c}{$^9$Be$^+$} & \multicolumn{3}{c}{$^{40}$Ar$^{15+}$ } \\
\cline{2-4} \cline{5-7} \cline{8-10} \\
State  & This work & Experiment & $\delta(\%)$ & This work & Experiment & $\delta(\%)$ & This work & Experiment & $\delta(\%)$ \\
\hline \\
$2s ~ ^2S_{1/2}$  & 43484(10) & 43487.1594(2) & 0.008 & 146876(20) &  146882.9(3)  & 0.005 & 7407(5)  & \textit{7407.3}(1) & 0.000 \\
$2p ~ ^2P_{1/2}$  & 28580(7) & 28583.5113(2) & 0.012 & 114940(15) &  114954.1(4)  & 0.012 & 7150(5)  & 7150.3(1) & 0.005 \\
$2p ~ ^2P_{3/2}$  & 28580(7) & 28583.1759(2) & 0.013 & 114933(15) &  114947.5(4)  & 0.013 & 7124(5)  & 7124.7(1) & 0.005 \\
$3s ~ ^2S_{1/2}$  & 16280(5) & 16281.064(1) & 0.005 &  58649(10) &   58650.9(5)  & 0.003 & 3232(3)  & 3231.3(5) & 0.010(17) \\
$3p ~ ^2P_{1/2}$  & 12561(2) & 12561.606(1) & 0.009 &  50383(7) &   50387.5(5)  & 0.009 & 3161(3)  & 3160.3(6) & 0.002(19) \\
$3p ~ ^2P_{3/2}$  & 12560(2) & 12561.510(1) & 0.009 &  50381(10) &   50385.6(5)  & 0.009 & 3153(3)  & 3153.1(7) & 0.008(21) \\
$3d ~ ^2D_{3/2}$  & 12204(1) & 12204.109(1) & 0.001 &  48828(5) &   48828.3(5)  & 0.001 & 3126(3)  & 3126.3(2) & 0.005(06) \\
$3d ~ ^2D_{5/2}$  & 12204(1) & 12204.073(1) & 0.001 &  48827(5) &   48827.8(5)  & 0.001 & 3124(3)  & 3123.7(8) & 0.001(26) \\
\end{tabular}
\end{ruledtabular}
\label{tabeng}
\end{table*} 

\begin{table}[t]
\caption{Comparison of excitation energies (in eV) of the D1 and D2 transitions of Ar$^{15+}$ from the present work with other high-precision calculations using few-body approaches.}
\begin{ruledtabular}
\begin{tabular}{lrr}
Transition  &  This work & Others \\
\hline \\
$2s ~ ^2S_{1/2} - 2p ~ ^2P_{1/2}$ & 31.8635 & 31.8681 \cite{Sapirstein}  \\
                                  &  & 31.8673(5) \cite{Kozhedub} \\  
$2s ~ ^2S_{1/2} - 2p ~ ^2P_{3/2}$ & 35.0870 & 35.0371 \cite{Sapirstein} \\
                                  &     &  35.0378(6) \cite{Kozhedub}   \\
\end{tabular}
\end{ruledtabular}
\label{tabeng1}
\end{table}

\section{Methods}

\subsection{RCC theory of one-valence systems}

In the RCC theory {\it ans\"atz}, the wave function of an atomic state of the considered systems is given by \cite{Lindgren,Mukherjee,Sahoo_2004}
\begin{eqnarray}
 |\Psi_v \rangle = e^T \{1+S_v \} |\Phi_v \rangle,
 \label{eqcc}
\end{eqnarray}
where $|\Phi_v \rangle$ is a mean-field wave function and is constructed in the present work as $|\Phi_v \rangle = a_v^{\dagger} |\Phi_0 \rangle$ with the Dirac-Hartree-Fock (DHF) wave function $|\Phi_0 \rangle$ of the closed-core $1s^2$, and $T$ and $S_v$ are the excitation operators that are responsible for accounting the electron correlation effects from the core orbitals and valence orbital, respectively. Since Li-like systems have only two core orbitals and one valence orbital, the exact form of the above expression yields
\begin{eqnarray}
 |\Psi_v \rangle = [(1+ T+ \frac{1}{2} T^2) \{1+S_v \}] |\Phi_v \rangle .
\end{eqnarray}

Following the Schr\"odinger equation $H |\Psi_v \rangle = E_v|\Psi_v \rangle$ with the atomic Hamiltonian $H$ and energy eigenvalue $E_v$ of the corresponding state, the $T$ and $S_v$ amplitude solving equations are given by
\begin{eqnarray}
 && \langle \Phi_0^* | \bar{H} | \Phi_0 \rangle = 0 \label{eqt0} \\
  \text{and} && \nonumber \\
 && \langle \Phi_v^* | \{ (\bar{H}-E_v) S_v \} + \bar{H} | \Phi_v \rangle = 0, \label{eqamp}
\end{eqnarray}
respectively, where the superscript $*$ over the reference states indicates that the states refer to the excited determinants with respect to the respective reference states, and $\bar{H}= e^{-T}He^T = (He^T)_l$ with the subscript $l$ denoting the linked terms. The energies of the states having closed-shell ($E_0$) and closed-shell with the valence orbital configurations are obtained by 
 \begin{eqnarray}
 && E_0=\langle \Phi_0 | \bar{H} | \Phi_0 \rangle \label{eqeng0}  \\
  \text{and} && \nonumber \\
&& E_v = \langle \Phi_v | \bar{H}\{ 1+ S_v \} | \Phi_v \rangle. \label{eqeng}
\end{eqnarray}
Since Eq. (\ref{eqamp}) depends on $E_v$, both Eq. (\ref{eqamp}) and Eq. (\ref{eqeng}) are solved simultaneously. Instead of calculating $E_0$ and $E_v$ separately, we evaluate the electron affinity of an electron from the valence orbital of the atomic state by taking the difference $\Delta E_v=E_v - E_0$, and is also equivalent to the ionization potential (I.P.) of the valence electron.

We have approximated the RCC excitation operators $T$ and $S_v$ at the singles, doubles and triples excitations, denoted by
\begin{eqnarray}
 T = T_1 + T_2 \ \ \ \text{and} \ \ \ S_v = S_{1v} + S_{2v} + S_{3v} ,
\end{eqnarray}
where the subscripts $1,2$ and $3$ denote for the level of excitation respectively. It should be noted that the $T$ operator can only have up to the double excitations as there are only two core electrons present in the Li-like atomic systems. To fathom contributions from the triple excitations arising through the $S_{3v}$ operator, we also compute calculations considering only the singles and doubles excitations in the $S_v$ operator by expressing $S_v = S_{1v} + S_{2v}$ in the RCC theory (denoted by RCCSD method).

\begin{table*}[t]
\caption{Comparison of the FS constants (in MHz fm$^{-2}$) from the FF, EVE and AR approaches using the DC Hamiltonian in Li, Be$^+$ and Ar$^{15+}$. Results are given from the DHF, RCCSD and full RCC calculations to demonstrate trends of electron correlations. The anomalous behaviour of results with the FF approach is discussed in the text.}
\begin{ruledtabular}
\begin{tabular}{l rrr rrr rrr}
  & \multicolumn{3}{c}{FF approach} & \multicolumn{3}{c}{EVE approach} & \multicolumn{3}{c}{AR approach} \\
\cline{2-4} \cline{5-7} \cline{8-10} \\
State  & DHF & RCCSD & RCC & DHF & RCCSD & RCC & DHF & RCCSD & RCC \\
\hline \\
\multicolumn{10}{c}{\underline{Li atom}} \\
$2s ~ ^2S_{1/2}$ & $-5.882$ & $-5.712$   & $-1.732$ &  $-2.428$ & $-2.043$ & $-2.038$  & $-2.428$ & $-2.037$  & $-2.035$ \\
$2p ~ ^2P_{1/2}$ & 0.795 & 1.139 & 1.138 & $-0.000$ & 0.415 & 0.421  & $-0.000$ & 0.424 & 0.427 \\
$2p ~ ^2P_{3/2}$ & 0.791 & 1.135 & 1.135 & $-0.000$ & 0.414 & 0.421  &  $-0.000$ & 0.424 & 0.426\\
$3s ~ ^2S_{1/2}$ & $-1.377$ &  $-1.336$  & $-0.703$ & $-0.571$ & $-0.479$ & $-0.474$ & $-0.571$  & $-0.478$  & $-0.477$ \\
$3p ~ ^2P_{1/2}$ & 0.258  & 0.352 & 0.352 &  $-0.000$ & 0.128 & 0.129 & $-0.000$ & 0.131  & 0.132 \\
$3p ~ ^2P_{3/2}$ & 0.257 & 0.350 & 0.350 & $-0.000$ & 0.128 & 0.128  &$-0.000$  & 0.131  & 0.131 \\
$3d ~ ^2D_{3/2}$ & $-0.004$ & 0.008  & 0.008 &$-0.000$ & 0.005 & 0.005  & $-0.000$ &  0.005  & 0.005 \\
$3d ~ ^2D_{5/2}$ & $-0.001$ & 0.010 & 0.010 & $-0.000$ & 0.005 & 0.005  & $-0.000$ & 0.005  & 0.005 \\
& & & \\
\multicolumn{10}{c}{\underline{Be$^+$ ion}} \\
$2s ~ ^2S_{1/2}$ & $-15.557$ & $-15.365$  & $-34.853$  &  $-15.712$ & $-14.029$ & $-13.974$   & $-15.712$  &  $-14.015$  &  $-13.991$\\
$2p ~ ^2P_{1/2}$ & 2.686  & 3.257  & 3.262 &  $-0.000$ & 2.982 & $3.001$  & $-0.000$ & 3.019 & 3.027 \\
$2p ~ ^2P_{3/2}$ & 2.689  & 3.260  & 3.256 &  $0.000$ & 2.981 & 3.002  &  $0.000$  & 3.017 & 3.025 \\
$3s ~ ^2S_{1/2}$ &  $-3.989$  & $-3.937$   & $-6.634$ & $-4.039$ & $-3.595$ &  $-3.597$  &  $-4.039$  & $-3.592$  & $-3.591$\\
$3p ~ ^2P_{1/2}$ & 0.809 & 0.945 & 0.947  &  $-0.000$ & 0.862 & 0.871  & $-0.000$ &  0.873  &  0.876  \\
$3p ~ ^2P_{3/2}$ & 0.812  & 0.947  & 0.946 &  $0.000$ & 0.862 & 0.870  & $0.000$  & 0.872  &  0.875 \\
$3d ~ ^2D_{3/2}$ &  0.008 & 0.041  & 0.041 &  $0.000$& 0.045 & 0.044      & $0.000$ & 0.046 & 0.045\\
$3d ~ ^2D_{5/2}$ &  0.008 & 0.041  & 0.041 &  $0.000$ & 0.045 & 0.044  &  $0.000$ & 0.046  & 0.045 \\
& & & \\
\multicolumn{10}{c}{\underline{Ar$^{15+}$ ion}} \\
$2s ~ ^2S_{1/2}$ & $-18819$  & $-18811$  & $-19154$  & $-19203$ & $-18820$ & $-18820$  &  $-19203$  &  $-18820$  & $-18820$ \\
$2p ~ ^2P_{1/2}$   & $936$  & $958$  &  958 &  $-53$  & 960  & 961           &  $-53$  &  961  &  961 \\
$2p ~ ^2P_{3/2}$ & 967  & 989  &  988 &  $-0$ & 989 & 989    &  $-0$  & 989  &  990 \\
$3s ~ ^2S_{1/2}$ & $-5447$  & $-5444$  & $-5574$  & $-5564$ & $-5447$ & $-5447$     &  $-5564$  &  $-5447$  & $-5447$\\
$3p ~ ^2P_{1/2}$ & $245$  &  $249$  & 249 & $-18$ & 250 & 250   &  $-18$   &  250  & 250  \\
$3p ~ ^2P_{3/2}$ & 257  & 262   & 262  &  $-0$ & 262 & 262  &  $-0$  &  262  &  262   \\
$3d ~ ^2D_{3/2}$ & 18 & 21  & 21 &  $-0$ & 21  & 21    & $-0$  &  21  &  21  \\
$3d ~ ^2D_{5/2}$ & 17 & 20 & 20  &  $-0$ & 21 & 21  & $-0$  &  21  &  21  \\
\end{tabular}
\end{ruledtabular}
\label{FScomp}
\end{table*}

\subsection{Evaluation of IS constants in the FF approach}

The IS constant of the respective $O$ operator can be determined in the FF approach by using an effective Hamiltonian $H=H_{0} + \lambda O$, where $H_{0}$ is the atomic Hamiltonian without the IS interactions and $\lambda$ is an arbitrary parameter. Then, the electron affinity obtained by considering the above Hamiltonian can be expressed as
\begin{eqnarray}
\Delta E_v(\lambda) &=& \Delta E_v^{(0)} + \lambda \Delta E_v^{(1)} + {\cal O}(\lambda)^{2} ,
\label{eqff}
\end{eqnarray}
where superscripts (0), (1), etc. denote the order of perturbation and ${\cal O}(\lambda)^2$ indicates corrections higher than the first-order. For a very small value of $\lambda$, we get
\begin{eqnarray}
\Delta E_v(\lambda) \approx \Delta E_v^{(0)} + \lambda \Delta E_v^{(1)}  .
\end{eqnarray}
Therefore, the first-order energy correction can be estimated from the above expression as
\begin{eqnarray}
 \Delta E_v^{(1)} \equiv \left. \frac{\partial \Delta E_v(\lambda)}{\partial \lambda} \right|_{\lambda=0}  \approx \frac{\Delta E_v(+\lambda)-\Delta E_v(-\lambda)}{2 \lambda}.
\label{eqff1}
\end{eqnarray}

In the perturbative approach, it corresponds to $\Delta E_v^{(1)} = \langle O \rangle$. This approach is commonly adopted in the evaluation of the IS constants. However, we would like to draw attention of several drawbacks of this approach. First of all, it forcefully assumes that ${\cal O}(\lambda)^{2}$ contributions in Eq. (\ref{eqff1}) are negligible compared with the first-order contribution. This may not always be the case, thus, it could result in numerical inaccuracy in the estimation. This inaccuracy could potentially be overcome by choosing $\lambda$ values differently for the FS, NMS and SMS constants, which is obvious from Eq. (\ref{IS}). Moreover, this could also be state dependent for a given atomic system. Therefore, appropriate choice of $\lambda$ for the respective IS constant requires knowledge of their strengths {\it a priori} which is not always possible. In this work, we have used a fixed value $\lambda=1.0\times 10^{-5}$ to determine the FS, NMS and SMS constants to carry out our analysis. Further smaller value of $\lambda$ may lead to truncation errors in the calculations.

\subsection{Evaluation of IS constants in the EVE approach}

It is obvious from the above discussion that the IS constants can also be evaluated directly in the EVE approach by using the wave functions of the unperturbed Hamiltonian $H_0$. In fact, any general physical property of an atomic system is evaluated as the expectation value of the corresponding operator. Thus, this should be the most convenient approach to evaluate the IS constants. However, EVE is less popular than the FF approach in the context of calculating the IS constants. This is owing to the fact that FF can account for the orbital relaxation effects at the DHF level, which are missing in the DHF values of the EVE approach. Since these orbital relaxation effects are quite strong due to the IS effects, they have to be adequately included in the post DHF calculations through the EVE approach. 

In the RCC theory framework, the EVE expression is given by 
\begin{eqnarray}
\langle O \rangle &=&  \frac{\langle \Psi_v | O | \Psi_v \rangle}{\langle \Psi_v |\Psi_v \rangle} \nonumber \\
&=& \frac{\langle \Phi_v | \{ 1+ S_v^{\dagger} \} e^{T^{\dagger}} O e^T \{1+S_v \} | \Phi_v \rangle} {\langle \Phi_v | \{ 1+ S_v^{\dagger} \} e^{T^{\dagger}} e^T \{1+S_v \} | \Phi_v \rangle}  ,
\label{evexp}
\end{eqnarray}
where amplitudes of the $T$ and $S_v$ operators are obtained using $H_0$ in Eqs. (\ref{eqt0}) and (\ref{eqamp}), respectively. As can be seen, the above expression contains two non-terminating series, namely, $e^{T^{\dagger}} O e^T$ and $e^{T^{\dagger}} e^T$ in the numerator and denominator respectively. Moreover, the above expression does not satisfy the Hellmann-Feynman theorem \cite{Bishop,Bishop1991}. These are the main disadvantages of adopting the EVE approach for the determination of IS constants in the RCC theory framework.

\begin{table*}[t]
\caption{The FS constants (in MHz fm$^{-2}$) from the DCB and DCQ Hamiltonians are given using the DHF, RCCSD and RCC calculations by adopting the AR approach. The RCC results adding DC, DCB and DCQ contributions (DCBQ) are given in the last column along with their uncertainties are given in the last column.}
\begin{ruledtabular}
\begin{tabular}{l rrr rrr c}
    & \multicolumn{3}{c}{DCB} & \multicolumn{3}{c}{DCQ} & DCBQ \\
\cline{2-4} \cline{5-7}  \\
State  & DHF  &  RCCSD  &  RCC &  DHF  &  RCCSD  &  RCC & RCC   \\
\hline \\
\multicolumn{8}{c}{\underline{Li atom}} \\
$2s ~ ^2S_{1/2}$  &  $-2.428$ & $-2.036$ & $-2.035$ &  $-2.423$  & $-2.033$  &  $-2.031$ & $-2.031(3)$ \\
$2p ~ ^2P_{1/2}$  &  $-0.000$ & 0.424 & 0.426 &$-0.000$ &  0.423 & 0.426 & 0.425(1) \\
$2p ~ ^2P_{3/2}$  &  $-0.000$ & 0.424 & 0.426 & $-0.000$ & 0.423 & 0.425 & 0.425(1) \\
$3s ~ ^2S_{1/2}$  &  $-0.571$ & $-0.478$  & $-0.476$ &   $-0.570$ &  $-0.477$  &  $-0.476$  & $-0.475(1)$ \\
$3p ~ ^2P_{1/2}$  &  $-0.000$& 0.131 &  0.131 &  $-0.000$ & 0.130  &  0.131 & 0.130(1) \\
$3p ~ ^2P_{3/2}$  &  $-0.000$ & 0.131 & 0.131  & $-0.000$ & 0.130  &  0.131 & 0.131(1)  \\
$3d ~ ^2D_{3/2}$  &  $-0.000$ & 0.005 &  0.005 & $-0.000$ & 0.005  &  0.005 & 0.005(0) \\
$3d ~ ^2D_{5/2}$  &  $-0.000$& 0.005 & 0.005  & $-0.000$ &  0.005 &  0.005  & 0.005(0)  \\
& & & \\
\multicolumn{8}{c}{\underline{Be$^+$ ion}} \\
$2s ~ ^2S_{1/2}$  & $-15.711$  &  $-14.012$ & $-13.992$  & $-15.677$ & $-13.983$  &  $-13.947$  & $-13.948(50)$ \\
$2p ~ ^2P_{1/2}$  &  $-0.000$  &  3.018 &  3.027  & $-0.000$   & 3.013  &  3.021 & 3.021(20)  \\
$2p ~ ^2P_{3/2}$  & $0.000$  & 3.016  &  3.024   & $0.000$    &  3.010 &  3.018 & 3.017(20) \\
$3s ~ ^2S_{1/2}$  & $-4.039$  &  $-3.591$  &  $-3.591$  &  $-4.030$  &  $-3.584$  &  $-3.582$ & $-3.582(30)$   \\
$3p ~ ^2P_{1/2}$  & $-0.000$   & 0.873  &  0.876  & $-0.000$  & 0.871  &  0.874 & 0.874(6) \\
$3p ~ ^2P_{3/2}$  & $0.000$ &  0.872 &  0.875  & $0.000$  & 0.870  &  0.873 & 0.873(6)  \\
$3d ~ ^2D_{3/2}$  & $0.000$ &  0.046 &  0.045  & $0.000$ & 0.045  &  0.045 & 0.045(1)  \\
$3d ~ ^2D_{5/2}$  & $0.000$ &  0.046 &  0.045  & $0.000$  & 0.045  &  0.045 & 0.045(1) \\
& & & \\
\multicolumn{8}{c}{\underline{Ar$^{15+}$ ion}} \\
$2s ~ ^2S_{1/2}$  &  $-19196$ & $-18799$  &  $-18799$  &  $-19035$ &  $-18655$ & $-18655$ & $-18633(50)$  \\
$2p ~ ^2P_{1/2}$  &  $-53$ & 959 &  959  &  $-52$  &  953  & 953  & 951(5) \\
$2p ~ ^2P_{3/2}$  &  $-0$  & 986 & 987 & $-0$  &  981  & 981 & 978(5)  \\
$3s ~ ^2S_{1/2}$  & $-5563$  & $-5442$ &  $-5442$  & $-5516$  &  $-5400$  &  $-5400$ & $-5394(10)$  \\
$3p ~ ^2P_{1/2}$  &  $-18$ & 248  & 248  & $-18$  &  248  & 248 & 236(3)   \\
$3p ~ ^2P_{3/2}$  & $-0$  &  261 & 261 & $-0$  &  260  & 259 & 258(3) \\
$3d ~ ^2D_{3/2}$  & $-0$  & 21 &  21 & $-0$  &  21 & 21 & 21.0(5)  \\
$3d ~ ^2D_{5/2}$  & $-0$  & 21 &  21 & $-0$  &  21  & 21 & 21.0(5)   \\
\end{tabular}
\end{ruledtabular}
\label{FSfin}
\end{table*} 

\begin{table*}[t]
\caption{Comparison of the NMS constants (in GHz amu) from the FF, EVE and AR approaches using the DC Hamiltonian in Li, Be$^+$ and Ar$^{15+}$. Results are given from the DHC, RCCSD and full RCC calculations for all the three approaches. Anomalous results are highlighted in bold and are discussed in the text.}
\begin{ruledtabular}
\begin{tabular}{l rrr rrr rrr}
  & \multicolumn{3}{c}{FF approach} & \multicolumn{3}{c}{EVE approach} & \multicolumn{3}{c}{AR approach} \\
\cline{2-4} \cline{5-7} \cline{8-10} \\
State  & DHF & RCCSD & RCC & DHF & RCCSD & RCC & DHF & RCCSD & RCC \\
\hline \\
\multicolumn{10}{c}{\underline{Li atom}} \\
$2s ~ ^2S_{1/2}$   & 708.426  & 714.900  & 716.249 &  747.093 & 713.654 & 713.303 & 747.093  & 712.965 & 713.031 \\
$2p ~ ^2P_{1/2}$   & 464.201  &  469.845 & 469.938 &  508.602 & \bf{474.621} & 473.741 &  508.602  &  468.472  & 468.051 \\
$2p ~ ^2P_{3/2}$   &  464.249  & 469.891 & 469.958 &  508.613 & 469.744 & 468.973 &  508.613  &  468.516  & 468.139 \\
$3s ~ ^2S_{1/2}$   & 266.250  & 267.609  & 267.566 & 275.655 & 267.447 & 267.130 &  275.655  &  267.303  & 267.164 \\
$3p ~ ^2P_{1/2}$    & 204.947  & 206.589  & 206.619  & 219.236 & 206.558 & 206.185 &  219.236  &  206.152  & 206.011 \\
$3p ~ ^2P_{3/2}$    & 204.844  & 206.477  & 206.499  & 219.239 & 206.483 & 205.934 &  219.239  &  206.166  &  206.040\\
$3d ~ ^2D_{3/2}$    & 200.578  &  200.750 & 200.752  & 200.651 & 200.665 & 200.663 &  200.651  &  200.638  &  200.642 \\
$3d ~ ^2D_{5/2}$   & 200.417  & 200.588  & 200.588 & 200.653 & 200.664 & 200.662   &  200.653  &  200.640  &  200.643  \\
& & & \\
\multicolumn{10}{c}{\underline{Be$^+$ ion}} \\
$2s ~ ^2S_{1/2}$  &  2404.41  & 2415.33  & \bf{1013.71} &  2506.87 & 2413.63 & 2411.34 &  2506.87  &  2412.65  &  2411.72 \\
$2p ~ ^2P_{1/2}$  & 1874.33  & 1889.57  & 1889.66&  2080.43 & \bf{1908.47} & 1906.45 &  2080.43  &  1887.23  &  1886.16 \\
$2p ~ ^2P_{3/2}$  & 1874.44  & 1889.66  & 1889.73 &  2080.39 & 1889.98 & 1888.20 &  2080.39  &  1887.38  &  1886.42 \\
$3s ~ ^2S_{1/2}$   & 962.03 & 964.58  & \bf{583.10} &  989.46 & 964.20 & 964.321   &   989.46  &  963.97  &  964.18 \\
$3p ~ ^2P_{1/2}$     & 824.39  & 828.47  & 828.50 &  885.61 & 828.72 & 827.96  &  885.61  &  827.88  &  827.54 \\
$3p ~ ^2P_{3/2}$     & 824.50   & 828.57  & 828.59 &  885.59 & 828.46 & 828.21  &  885.59  &  827.92   &  827.62 \\
$3d ~ ^2D_{3/2}$   & 802.48   & 803.15  & 803.15 &  803.53 & 802.90 & 802.91   &  803.52  &  802.81   &  802.83 \\
$3d ~ ^2D_{5/2}$     & 802.19   & 802.85  & 802.85 &  803.53 & 802.89 & 802.90  &  803.53  &  802.82  &  802.83 \\
& & & \\
\multicolumn{10}{c}{\underline{Ar$^{15+}$ ion}} \\
$2s ~ ^2S_{1/2}$   & 121634 & 121649 & 121691  & 122796  & 121656 &  121655 &  122796  &  121655  & 121655 \\
$2p ~ ^2P_{1/2}$   & 117356 & 117393 &  117394 & 120795  & 117465 & 117464 &  120795  &  117391  &  117390  \\
$2p ~ ^2P_{3/2}$   & 117006 & 117041   & 117041 & 120368  & 117039 & 117037 &  120368  &  117038  &  117037 \\
$3s ~ ^2S_{1/2}$      & 53094 & 53099   & 53117 & 53456 & 53100 &  53100 &  53456   &  53100 &  53100 \\
$3p ~ ^2P_{1/2}$       & 51908 &  51917 &  51917 & 52796 & 51917 & 51917 & 52796 &  51917  &  51916 \\
$3p ~ ^2P_{3/2}$      & 51803 & 51811  &  51811& 52672 & 51810 & 51810   &  52672  &  51810  &  51810  \\
$3d ~ ^2D_{3/2}$     & 51369 & 51371 &  51371 & 51439 & 51371& 51371   & 51439 &  51371  &  51371 \\
$3d ~ ^2D_{5/2}$    & 51340 & 51342   &  51342 & 51409 & 51342& 51342     &  51409 & 51342  &  51342 \\
\end{tabular}
\end{ruledtabular}
\label{NMScomp}
\end{table*}

\subsection{Evaluation of IS constants in the AR approach}

The problems of the FF and EVE approaches in the determination of the IS constants can be circumvented by applying the AR approach in the RCC theory framework \cite{Monkhorst}. The basic notion of this approach lies in the fact that it evaluates the IS constants as the first-order energy correction like in the FF approach, {\it albeit} its starting point is the same as in the EVE approach. However, unlike FF, AR does not depend on the perturbative parameter $\lambda$, and higher-order perturbative corrections do not appear in the expression. 

Following the FF approach, the wave function and energy due to the total Hamiltonian $H=H_{0} + \lambda O$ can be expanded as
\begin{eqnarray}
&& |\Psi_v \rangle = |\Psi_v^{(0)} \rangle + \lambda  |\Psi_v^{(1)}\rangle + {\cal O}(\lambda)^{2}, \nonumber \\
&& E_0 = E_0^{(0)} + \lambda E_0^{(1)} + {\cal O}(\lambda)^{2} \nonumber \\ 
\text{and} && \nonumber \\
&& E_v = E_v^{(0)} + \lambda E_v^{(1)} + {\cal O}(\lambda)^{2} ,
\end{eqnarray}
where superscripts denote order of perturbation. Different orders of wave functions can be obtained by expanding the RCC operators as
\begin{eqnarray}
 && T= T^{(0)} + \lambda T^{(1)}+ {\cal O}(\lambda)^{2} \\
 \text{and} && \nonumber \\
 && S_{v}= S_{v}^{(0)} + \lambda S_v^{(1)}+ {\cal O}(\lambda)^{2} .
\end{eqnarray}
Substituting the expanded forms of the Hamiltonian and RCC operators in Eqs. (\ref{eqeng0}) and (\ref{eqeng}), and retaining only the terms linear in $\lambda$, we get 
\begin{eqnarray}
&& E_0^{(1)} = \langle \Phi_0 | \overline{H_0} T^{(1)} + \overline{O} | \Phi_0 \rangle \label{eqeng00} \\
\text{and} && \nonumber \\
&& E_v^{(1)} = \langle \Phi_v | \overline{H_0} S_v^{^{(1)}}  + (\overline{H_0} T^{(1)} + \overline{O}) \{ 1 + S_v^{^{(0)}} \} | \Phi_v \rangle ,  \label{eqeng1}
\end{eqnarray}
where $\overline{O}=(Oe^{T^{(0)}})_l$. Therefore, we can estimate the first-order correction to electron affinity due to $O$ by 
\begin{eqnarray}
 \Delta E_v^{(1)} = E_v^{(1)}- E_0^{(1)} \equiv \langle O \rangle .
\end{eqnarray}
It is evident from these equations that the above procedure of evaluating IS constants does not depend on the choice of $\lambda$, and that the lowest-order contributions are the same as the values obtained in the EVE approach. The amplitudes of the $T^{(1)}$ and $S_v^{(1)}$ operators are obtained by solving the following equations
\begin{eqnarray}
 && \langle \Phi_0^* | \overline{H}_0T^{(1)} + \overline{O}  | \Phi_0 \rangle = 0 \label{eqt1} \\
  \text{and} && \nonumber \\
 && \langle \Phi_v^* | \{ (\overline{H}_0-E_v^{(0)}) S_v^{(1)}+ \} +  \left (\overline{H}_0 T^{(1)} \right. \nonumber \\
 &&  \left. \ \ \ \ \ \ \ \ \  \ \ \ \ \ \ \ \ \ \ \ \ \ +\overline{O} \right ) \{1+ S_v^{(0)}| \Phi_v \rangle = 0. \label{eqamp1}
\end{eqnarray}
The AR approach, however, is more computationally demanding due to evaluation of amplitudes of both the unperturbed and perturbed RCC operators. 

\begin{table*}[t]
\caption{The NMS constants (in GHz amu) from the DCB and DCQ Hamiltonians using the DHF, RCCSD and RCC calculations by considering the AR approach. Combined DCBQ results along with uncertainties and the values obtained from the scaling law are also given.}
\begin{ruledtabular}
\begin{tabular}{l rrr rrr cr}
     & \multicolumn{3}{c}{DCB} & \multicolumn{3}{c}{DCQ} & DCBQ & Scaling \\
\cline{2-4} \cline{5-7}  \\
State  & DHF  &  RCCSD  &  RCC &  DHF  &  RCCSD  &  RCC & RCC &  energy  \\
\hline \\
\multicolumn{9}{c}{\underline{Li atom}} \\
$2s ~ ^2S_{1/2}$  &  747.070  & 712.786  &  712.844  &  747.075  &  712.808  &  712.873 & 713(2) & 715.195  \\
$2p ~ ^2P_{1/2}$  &  508.564  & 468.365  &  467.942  &  508.605  & 468.400  &  467.869  & 468(1) & 470.052  \\
$2p ~ ^2P_{3/2}$  &  508.601  & 468.432  &  468.053  &  508.616  & 468.445  &  467.981  & 468(1) & 470.045  \\
$3s ~ ^2S_{1/2}$  &  275.649  & 267.263  &  267.121  &  275.650  &  267.268  &  267.127 & 267.1(5) & 267.766  \\
$3p ~ ^2P_{1/2}$  &  219.223  & 206.119  &  205.977  &  219.237  &  206.130 &  205.989  & 206.0(8) & 206.582  \\
$3p ~ ^2P_{3/2}$  &  219.234  & 206.140  &  206.013  &  219.240  &  206.144 &  206.018  & 206.0(8) & 206.580  \\
$3d ~ ^2D_{3/2}$  &  200.651  & 200.638  &  200.641  &  200.651  & 200.638  &  200.641  & 200.64(5) & 200.723  \\
$3d ~ ^2D_{5/2}$  &  200.653  & 200.639  &  200.643  &  200.653  & 200.639  &  200.643  & 200.64(5) & 200.722  \\
& & & \\
\multicolumn{9}{c}{\underline{Be$^+$ ion}} \\
$2s ~ ^2S_{1/2}$  &  2506.75  & 2412.46  &  2411.79  &  2506.78  &  2412.55  &  2410.80  & 2411(4) & 2415.67  \\
$2p ~ ^2P_{1/2}$  &  2080.08  & 1886.92  &  1885.85  &  2080.45  & 1887.25   &  1886.18 & 1886(3) & 1890.37 \\
$2p ~ ^2P_{3/2}$  & 2080.25  & 1887.26  &  1886.30  &  2080.40  &  1887.39  &  1886.44  & 1886(3) & 1890.25 \\
$3s ~ ^2S_{1/2}$  & 989.44  &  963.92  &   964.18  & 989.44  & 963.94  &   964.07 & 964(2) & 964.60 \\
$3p ~ ^2P_{1/2}$  &  885.51  &  827.79  &   827.46  &  885.61  & 827.88  &  827.55 & 828(2) & 828.63  \\
$3p ~ ^2P_{3/2}$  & 885.56  &  827.89  &   827.59  &  885.60  & 827.92  &  827.63 & 828(2) & 828.60 \\
$3d ~ ^2D_{3/2}$  &  803.52  &  802.81  &   802.83  &  803.53  & 802.81 &  802.83 & 803(1) & 803.07  \\
$3d ~ ^2D_{5/2}$  & 803.53  &  802.82  &   802.84  &  803.53  & 802.82  &  802.83 & 803(1) & 803.06\\
& & & \\
\multicolumn{9}{c}{\underline{Ar$^{15+}$ ion}} \\
$2s ~ ^2S_{1/2}$  & 122766 &  121591 & 121590  &  122754  & 121614  &  121613  & 121549(300) & 121824 \\
$2p ~ ^2P_{1/2}$  &  120663 & 117267  & 117266 &  120795  &  117391 &  117390  & 117267(250) & 117588 \\
$2p ~ ^2P_{3/2}$  & 120311 & 116991  & 116989  & 120369 &  117039 &  117039 & 116990(250) &  117168 \\
$3s ~ ^2S_{1/2}$  & 53448  &  53082 &  53082  & 53444  &  53088  &   53088  & 53070(100) & 53148\\
$3p ~ ^2P_{1/2}$  & 52763  & 51890  &  51889  & 52796  &  51917 &   51916 & 51890(100) & 51976 \\
$3p ~ ^2P_{3/2}$  & 52659   &  51802&  51801  & 52672 & 51811 &   51811 & 51802(100) & 51853 \\
$3d ~ ^2D_{3/2}$  &  51437   &  51368 & 51368   &  51439  &  51371 &   51371 & 51368(70) & 51414 \\
$3d ~ ^2D_{5/2}$  & 51408  &  51341 &  51341  & 51409 &  51342 &   51342  & 51341(70) &  51375\\
\end{tabular}
\end{ruledtabular}
\label{NMSfin}
\end{table*}

\subsection{Approximations in $H_0$}

The first approximation in our calculation is taken in the atomic Hamiltonian $H_0$, which is considered initially as the Dirac-Coulomb (DC) Hamiltonian. The DC Hamiltonian in atomic units (a.u.) is given by
\begin{eqnarray}\label{eq:DC}
H^{DC} &=& \sum_i \left [c\mbox{\boldmath$\alpha$}_i^D \cdot \textbf{p}_i+(\beta_i-1)c^2+V_n(r_i)\right] +\sum_{i,j>i}\frac{1}{r_{ij}}, \ \ \ \
\end{eqnarray}
where $c$ is speed of light, $\beta$ is the Dirac matrix, $\textbf{p}$ is the single particle momentum operator and $\sum_{i,j}\frac{1}{r_{ij}}$ represents the Coulomb potential between the electrons located at the $i^{th}$ and $j^{th}$ positions. It should be noted that the above Hamiltonian is scaled with respect to the rest mass energies of electrons and calculations are performed by using the electron mass in the kinetic energy term rather than the reduced mass $\mu_A$. The notation for the dependency of $V_n(r)$ on the rms radius of the nucleus is dropped for the convenience. Corrections to the energies due to finite mass of the nucleus are included separately at the later stage. Contributions from the Breit interaction to the DC Hamiltonian (DCB Hamiltonian) are determined by including the following potential 
\begin{eqnarray}\label{eq:DHB}
V^B &=& - \sum_{j>i}\frac{[\mbox{\boldmath$\alpha$}_i^D \cdot\mbox{\boldmath$\alpha$}_j^D +
(\mbox{\boldmath$\alpha$}_i^D \cdot\mathbf{\hat{r}_{ij}})(\mbox{\boldmath$\alpha$}_j^D \cdot \mathbf{\hat{r}_{ij}})]}{2r_{ij}} ,
\end{eqnarray}
where $\mathbf{\hat{r}_{ij}}$ is the unit vector along $\mathbf{r_{ij}}$.

\begin{table*}[t]
\caption{Comparison of the SMS constants (in GHz amu) from the FF, EVE and AR approaches using the DC Hamiltonian in Li, Be$^+$ and Ar$^{15+}$. Results are given from the DHC, RCCSD and full RCC calculations for all the three approaches. Anomalous values are highlighted in bold and are discussed in the text.}
\begin{ruledtabular}
\begin{tabular}{l rrr rrr rrr}
  & \multicolumn{3}{c}{FF approach} & \multicolumn{3}{c}{EVE approach} & \multicolumn{3}{c}{AR approach} \\
\cline{2-4} \cline{5-7} \cline{8-10} \\
State  & DHF & RCCSD & RCC & DHF & RCCSD & RCC & DHF & RCCSD & RCC \\
\hline \\
\multicolumn{10}{c}{\underline{Li atom}} \\
$2s ~ ^2S_{1/2}$    &  0.000    &  45.472  & 46.319 &  0.000  & 42.818 &  43.435           &  0.000  &  44.489  & 45.243    \\
$2p ~ ^2P_{1/2}$   &  $-150.156$    & $-150.562$   & $-152.842$  & $-150.220$ & $\bf{-158.809}$ & $-160.301$ &  $-150.220$  & $-150.737$  & $-152.454$ \\
$2p ~ ^2P_{3/2}$   &  $-150.239$   &  $-150.644$  & $-152.531$  &  $-150.219$ & $\bf{-158.966}$ & $-160.461$ &  $-150.219$  &  $-150.736$  & $-152.478$ \\
$3s ~ ^2S_{1/2}$    &  0.000   &  10.801  & 11.041  &  0.000 & 10.175 & 10.177              &  0.000  & 10.568  &  10.753 \\
$3p ~ ^2P_{1/2}$   &  $-48.168$   &  $-46.772$  & $-47.293$  &  $-48.145$ & $\bf{-49.289}$ & $-49.581$   &  $-48.145$  & $-46.781$  & $-47.244$\\
$3p ~ ^2P_{3/2}$  &  $-48.104$   &  $-46.710$  & $-47.223$  & $-48.146$ & $\bf{-49.342}$ & $-49.494$     & $-48.146$   &  $-46.782$  &  $-47.252$\\
$3d ~ ^2D_{3/2}$   &  0.000  &  $-0.144$  & $-0.186$  &  0.000 & $-0.202$ & $-0.268$           &  0.000  &  $-0.148$   &  $-0.196$\\
$3d ~ ^2D_{5/2}$   &  0.000  &  $-0.144$  & $-0.186$  &  0.000 & $-0.203$ & $-0.268$           &  0.000  &  $-0.148$  & $-0.196$ \\
& & & \\
\multicolumn{10}{c}{\underline{Be$^+$ ion}} \\
$2s ~ ^2S_{1/2}$   &  0.00  &  115.52  & \bf{153.31} & 0.00 & 112.16 & 112.64               & 0.00  & 113.80  & 114.74 \\
$2p ~ ^2P_{1/2}$  & $-951.02$  &  $-913.11$  & $-917.96$  & $-950.73$ & $\bf{-936.97}$ & $-940.14$  &  $-950.73$  &  $-913.01$  & $-917.36$\\
$2p ~ ^2P_{3/2}$  &  $-950.86$ &  $-913.03$  & $-917.84$  & $-950.67$ & $\bf{-937.53}$  & $-940.56$ &  $-950.67$  &  $-913.01$  &  $-917.44$\\
$3s ~ ^2S_{1/2}$   & 0.00  & 30.38   &  \bf{32.05} &  0.00 & 29.40 & 29.90               &  0.000  &  29.92   &  30.20 \\
$3p ~ ^2P_{1/2}$  &  $-276.08$  &  $-255.53$  &  $-256.65$  & $-275.96$  & $\bf{-262.30}$ & $-262.89$  &  $-275.96$  &  $-255.45$  &  $-256.45$\\
$3p ~ ^2P_{3/2}$  &  $-275.97$  & $-255.44$   &  $-256.54$ & $-275.97$ & $\bf{-262.50}$ & $-263.19$  &  $-275.97$  &  $-255.49$  &  $-256.50$\\
$3d ~ ^2D_{3/2}$   & 0.00 &  $-0.73$  &  $-0.91$ & 0.00 &    $-0.94$ & $-1.20$           &  0.00  &  $-0.74$  &  $-0.94$ \\
$3d ~ ^2D_{5/2}$   & 0.00 &  $-0.73$  &  $-0.91$ & 0.00  & $-0.94$ & $-1.20$          &  0.00  &  $-0.74$  &  $-0.94$ \\
& & & \\
\multicolumn{10}{c}{\underline{Ar$^{15+}$ ion}} \\
$2s ~ ^2S_{1/2}$   & 0 &  1193 &  \bf{1237} &0& 1183 & 1187    & 0 & 1189  & 1192 \\
$2p ~ ^2P_{1/2}$   & $-73703$ & $-72155$  &  $-72167$ & $-73703$ &  $-72237$ & $-72242$  & $-73703$  &  $-72155$  &  $-72167$\\
$2p ~ ^2P_{3/2}$   &  $-73431$ & $-71927$   & $-71940$ & $-73431$ & $-72009$ & $-72014$    &  $-73431$  & $-71927$  & $-71939$ \\
$3s ~ ^2S_{1/2}$    & 0 &  360 & \bf{379}  & 0  & 356 & 358    & 0  &  359  & 360 \\
$3p ~ ^2P_{1/2}$   & $-17348$ &  $-16762$  & $-16765$  & $-17348$ & $-16783$ & $-16783$   &  $-17348$  & $-16762$ & $-16765$\\
$3p ~ ^2P_{3/2}$   & $-17405$ & $-16833$ &  $-16835$ & $-17405$ & $-16854$ & $-16854$   &  $-17405$  &  $-16833$ & $-16835$\\
$3d ~ ^2D_{3/2}$   & 0 & $-47$  & $-48$  & 0 & $-49$ & $-50$        &  0  &  $-47$  &  $-48$\\
$3d ~ ^2D_{5/2}$   & 0 &  $-47$ &  $-48$ &  0  & $-48$ & $-49$      &  0  &  $-47$  &  $-48$\\
\end{tabular}
\end{ruledtabular}
\label{SMScomp}
\end{table*}

Contributions from the QED effects to the DC Hamiltonian (DCQ Hamiltonian) are estimated by considering the lower-order vacuum polarization (VP) and the self-energy (SE) interactions. We account for $V_{VP}$ through the Uehling \cite{Uhlpot} and Wichmann-Kroll \cite{WKpot} potentials ($V_{VP}=V^{Uehl} + V^{WK}$), given by
\begin{eqnarray}
 \label{eq:uehl}
V^{Uehl}&=&- \frac{2}{3} \sum_i \frac{\alpha^2 }{r_i} \int_0^{\infty} dx \ x \ \rho(x)\int_1^{\infty}dt \sqrt{t^2-1} \nonumber \\
&& \times\left(\frac{1}{t^3}+\frac{1}{2t^5}\right)  \left [ e^{-2ct|r_i-x|} - e^{-2ct(r_i+x)} \right ] \ \ 
\end{eqnarray}
and
\begin{eqnarray}
 V^{WK} = \sum_i \frac{0.368 Z^2}{9 \pi c^3 (1+(1.62 c r_i )^4) } \rho(r_i),
\end{eqnarray}
respectively.

\begin{table*}[t]
\caption{The SMS constants (in GHz amu) from the DCB and DCQ Hamiltonians using the AR approach. The results from the DHF, RCCSD and RCC calculations are given from both the Hamiltonians. The last column presents DCBQ values along with the uncertainties.}
\begin{ruledtabular}
\begin{tabular}{l rrr rrr c}
     & \multicolumn{3}{c}{DCB} & \multicolumn{3}{c}{DCQ} & \multicolumn{1}{c}{DCBQ} \\
  \cline{2-4}  \cline{5-7} \\
State  & DHF  &  RCCSD  &  RCC &  DHF  &  RCCSD  &  RCC & RCC \\
\hline \\
\multicolumn{8}{c}{\underline{Li atom}} \\
$2s ~ ^2S_{1/2}$  &   0.000  & 44.537  &  45.371  & 0.000  & 44.490 &  45.244   &  45.4(5)   \\
$2p ~ ^2P_{1/2}$  &  $-150.223$  & $-150.752$ & $-152.469$  & $-150.225$ &  $-150.743$ &  $-152.460$ & $-153(1)$  \\
$2p ~ ^2P_{3/2}$  &  $-150.224$  & $-150.739$ &  $-152.480$  & $-150.225$  &  $-150.743$ & $-152.484$ & $-153(1)$  \\
$3s ~ ^2S_{1/2}$  &   0.000  & 10.583  &  10.814  &  0.000  &  10.568 &  10.753  & 10.8(2)  \\
$3p ~ ^2P_{1/2}$  &  $-48.146$  & $-46.785$  &  $-47.247$  & $-48.147$  &  $-46.783$ &  $-47.246$ & $-47.3(5)$  \\
$3p ~ ^2P_{3/2}$  &  $-48.147$  & $-46.782$  &  $-47.252$  & $-48.147$  &  $-46.784$ &  $-47.254$ & $-47.3(5)$  \\
$3d ~ ^2D_{3/2}$  &    0.000  & $-0.148$  &  $-0.196$  &  0.000  &  $-0.148$ &  $-0.195$ & $-0.195(1)$  \\
$3d ~ ^2D_{5/2}$  &    0.000  & $-0.148$  &  $-0.196$  &  0.000  & $-0.148$  & $-0.195$ & $-0.195(1)$  \\
& & & \\
\multicolumn{8}{c}{\underline{Be$^+$ ion}} \\
$2s ~ ^2S_{1/2}$  &  0.00  & 114.06  &  114.91  &  0.00  &  113.79 &  114.58  & 114.7(8) \\
$2p ~ ^2P_{1/2}$  &  $-950.72$  & $-912.99$  &  $-917.34$  & $-950.77$  & $-913.04$  &  $-917.39$ &  $-917(3)$  \\
$2p ~ ^2P_{3/2}$  & $-950.67$  & $-912.96$  &  $-917.38$  & $-950.70$  & $-913.03$  &  $-917.46$ &  $-917(3)$  \\
$3s ~ ^2S_{1/2}$  &  0.00  & 30.01  & 30.29   &  0.00  & 29.91  & 30.23 &  30.4(2) \\
$3p ~ ^2P_{1/2}$  &  $-275.97$  & $-255.44$  &  $-256.44$  & $-275.97$  &  $-255.46$ &  $-256.46$  &  $-257(2)$  \\
$3p ~ ^2P_{3/2}$  & $-275.97$  &  $-255.47$ &  $-256.48$  & $-275.98$  & $-255.49$  &  $-256.50$ &  $-257(2)$  \\
$3d ~ ^2D_{3/2}$  &  0.00   &  $-0.75$   &  $-0.94$  &  0.00  & $-0.74$  &  $-0.94$ &  $-0.94(3)$  \\
$3d ~ ^2D_{5/2}$  &  0.00   &  $-0.74$  &  $-0.94$  &  0.00  & $-0.74$  & $-0.94$ &  $-0.94(3)$ \\
& & & \\
\multicolumn{8}{c}{\underline{Ar$^{15+}$ ion}} \\
$2s ~ ^2S_{1/2}$  & 0   & 1258  &  1260  &  0  &  1190 &  1192 & 1260(10) \\
$2p ~ ^2P_{1/2}$  &  $-73700$  &  $-72128$ & $-72141$ & $-73710$ & $-72161$  &  $-72174$ & $-72147(200)$  \\
$2p ~ ^2P_{3/2}$  &  $-73418$  &  $-71892$ & $-71905$ & $-73439$  &  $-71934$  &  $-71946$ & $-71911(200)$ \\
$3s ~ ^2S_{1/2}$  & 0  &  387 &  388  &  0  & 359 & 360  & 388(5)  \\
$3p ~ ^2P_{1/2}$  & $-17357$  & $-16761$ &  $-16763$  & $-17347$ &  $-16762$  &  $-16764$ & $-16763(50)$ \\
$3p ~ ^2P_{3/2}$  & $-17404$  &  $-16825$ &  $-16827$  & $-17404$ & $-16832$  &  $-16835$ & $-16826(50)$ \\
$3d ~ ^2D_{3/2}$  & 0  &  $-49$ &  $-50$  &  0  & $-47$  &  $-48$ & $-50(1)$  \\
$3d ~ ^2D_{5/2}$  & 0  & $-47$  &  $-48$  &  0  &  $-47$  &  $-48$ & $-48(1)$  \\
\end{tabular}
\end{ruledtabular}
\label{SMSfin}
\end{table*}

The interaction potential due to SE ($V_{SE}$) is estimated by including two parts \cite{Flam05,Ginges16}
\begin{eqnarray}
V_{SE}^{ef}&=&  A_l \sum_i \frac{2 \pi Z \alpha^3 }{r_i} I_1^{ef}(r_i) - B_l \sum_i \frac{\alpha }{ r_i} I_2^{ef}(r_i) \ \ \
\end{eqnarray}
known as the effective electric form factor part and
\begin{eqnarray}
V_{SE}^{mg}&=& - \sum_k \frac{i\alpha^3}{4} \mbox{\boldmath$\gamma$} \cdot \mbox{\boldmath$\nabla$}_k \frac{1}{r_k} \int_0^{\infty} dx \ x \ \rho_n(x)
\int_1^{\infty} dt \frac{1}{t^3 \sqrt{t^2-1}} \nonumber \\
\times && \left [ e^{-2ct|r_k-x|} - e^{-2ct(r_k+x)} - 2ct \left (r_k+x-|r_k-x| \right ) \right ], \nonumber \\
\end{eqnarray}
known as the effective magnetic form factor part. In the above expressions, we use 
\begin{eqnarray}
A_l= \begin{cases} 0.074+0.35Z \alpha \ \text{for} \ l=0,1 \\  0.056+0.05 Z \alpha + 0.195 Z^2 \alpha_e \ \text{for} \ l=2  , \end{cases}
\end{eqnarray}
and
\begin{eqnarray}
B_l = \begin{cases} 1.071-1.97y^2 -2.128 y^3+0.169 y^4  \ \text{for} \ l=0,1 \\
     0 \ \text{for} \ l \ge 2 .  \end{cases}   
\end{eqnarray}
The integrals are given by
\begin{eqnarray}
I_1^{ef}(r) =  \int_0^{\infty} dx \ x \ \rho_n(x) [ (Z |r-x|+1) e^{-Z|r-x|} \nonumber \\  - (Z(r+x)+1) e^{-2ct(r+x)}  ] \ \ \ \ \ \
\end{eqnarray}
and
\begin{eqnarray}
 I_2^{ef}(r) &=& \int_0^{\infty} dx \ x \ \rho_n(x)  \int^{\infty}_1 dt \frac{1}{\sqrt{t^2-1}} \bigg \{ \left( 1-\frac{1}{2t^2} \right ) \nonumber \\
&\times& \left [ \ln(t^2-1)+4 \ln \left ( \frac{1}{Z \alpha_e} +\frac{1}{2} \right ) \right ]-\frac{3}{2}+\frac{1}{t^2} \big \}\nonumber \\
&\times& \{ \frac{\alpha_e}{t} \left [ e^{-2ct|r-x|} - e^{-2ct(r+x)} \right ] +2 r_A e^{2 r_A ct } \nonumber \\
&\times& \left [ E_1 (2ct (|r-x|+r_A)) - E_1 (2ct (r+x+r_A)) \right ] \bigg \} \nonumber \\
\end{eqnarray}
with the orbital quantum number $l$ of the system, $y=(Z-80)\alpha$, $r_A= 0.07 Z^2 \alpha^3$, and the exponential integral $E_1(r) = \int_r^{\infty} ds e^{-s}/s$.

We have derived the FS operator, nuclear potential and nuclear density by assuming a Fermi-charge distribution given by
\begin{equation}
\rho_{n}(r)=\frac{\rho_{0}}{1+e^{(r-b)/a}} 
\end{equation}
for the normalization factor $\rho_0$, the half-density radius $b$ fm and $a= 2.3/4(\text{ln}3)$ is related to the skin thickness. For the calculation of atomic wave function, we have used the relations
\begin{eqnarray}
&&b=\sqrt{\frac{5}{3} \langle r_{nuc}^2 \rangle - \frac{7}{3} \pi^2 a^2} \\
\text{and} && \nonumber \\
&& \langle r_{nuc}^2 \rangle^{1/2}  =0.836 M_A^{1/3} +0.57 
\label{eq:rapp}
\end{eqnarray}
in fm. We emphasize that the approximation made in Eq. (\ref{eq:rapp}) does not affect to the calculated $F^\text{FS}$ considerably, as the dependence of $F^{\text{FS}}$ on the absolute rms radius is small \cite{2010-Puch}.

\begin{table*}[t]
\caption{Comparison of FS constants in MHz/fm$^2$ determined in this work, with other theories. The comparison is made on the Breit level with the higher orders given separately.}
\begin{ruledtabular}
\begin{tabular}{l ccc c}
Transition  & EVE-RCC & AR-RCC  &  Others  &  Reference \\
\hline \\
\multicolumn{5}{c}{\underline{$^{6,7}$Li} atom} \\
2S-2P (DCB)  & $-2.459
(3)$ &  $-2.461(3)$ &     $-2.457$      & \cite{2008-Pach}\\
\quad QED correction   & &                  $0.005~$ &          ~ ~ $-0.009(2)$     & \cite{2013-Pach}   \\\\
2S-3S (DCB)  & $-1.565(3)$ &  $-1.559(3)$ &     $-1.566$         & \cite{2008-Pach}\\
\quad QED correction   & &                  $0.003~$ &   ~ ~ $-0.006(2)$     & \cite{2008-Pach}   \\\\
2S I.P (DCB) & $-2.038(3)$ &  $-2.035(3)$ &     $-2.046$        & \cite{2008-Pach}\\
\quad QED correction   & &                  $0.004~$ &  ~ ~ $-0.007(2)$    &  \cite{2010-Puch}\\

\\
\multicolumn{5}{c}{\underline{$^{9,10}$Be$^+$} ion} \\
2S-2P (DCB) & $-16.98(5)$ &  $-17.02(5)$ &  $-16.91$    & \cite{2008-Pach}\\
\quad QED correction   & &                  $0.05$ & ~ ~~ $-0.12(3)$     & \cite{2010-Puch}   \\\\

2S-3S (DCB) & $-10.38(6)$ &  $-10.37(6)$ &  $-10.38$  & \cite{2008-Pach}\\
\quad QED correction   & &                  $0.04$ &   ~ ~~ $-0.07(2)$   & \cite{2010-Puch}  \\\\
2S I.P(DCB) & $-13.98(5)$ &  $-13.99(5)$ &   $-13.95$ & \cite{2008-Pach}\\
\quad QED correction   & &                  $0.04$ &       ~ ~~ $-0.10(2)$   & \cite{2010-Puch}   \\\\

\multicolumn{5}{c}{\underline{Ar$^{15^+}$} ion} \\
2S-2P$_{1/2}$ (DCB)    & $-19757(50)$ & $-19758(50)$    & $-19751$    & \cite{2014-Sto}\\
\quad QED correction & &   $174 $  & ~ ~ ~ ~ ~  $44(2)$    & \cite{2014-Sto}\\\\
2S-2P$_{3/2}$ (DCB)    & $-19784(50)$ & $-19785(50)$    & $-19781$    & \cite{2014-Sto}\\
\quad QED correction & &   $174 $  & ~ ~ ~ ~ ~  $44(2)$    & \cite{2014-Sto}\\
\end{tabular}
\end{ruledtabular}
\label{FSanl}
\end{table*} 

\subsection{Basis functions}

We have used Gaussian type orbitals (GTOs) \cite{Boys} to construct the single particle DHF wave functions. The radial component of a DHF orbital from the orbital angular momentum $l$ are given by using these GTOs as
\begin{eqnarray}
 f(r) =  r^l \sum_k^{N_l} e^{-\eta \zeta^k r^2} ,
\end{eqnarray}
where $\eta$ and $\zeta$ are the optimized GTO parameters for a given orbital, and $N_l$ represents the number of GTOs for the corresponding $l$-symmetry orbital. We have considered 40 GTOs, and universal basis functions by considering $\eta=0.0009$ and $\zeta=2.15$ for each symmetry up to $l=4$. Numerical radial integration are carried out on a non-linear grid distribution, $i=1,n$ with number of grids $n$, by defining radial distance as
\begin{eqnarray}
r(i) = r_0 \left [e^{(i-1) h} - 1 \right ] ,
\end{eqnarray}
where $r_0$ is a very small parameter and $h$ is the step size. In our calculation, we have used $n=1200$ for all the atomic systems while we have considered $h=0.018$ and $r_0= 3.3 \times 10^{-7}$ for the Li atom, $h=0.018$ and $r_0= 2.5 \times 10^{-7}$ for the Be$^+$ ion, and $h=0.019$ and $r_0= 5.6 \times 10^{-8}$ for the Ar$^{15+}$ ion.

We have estimated uncertainties to all the calculated quantities from the use of the finite-size basis functions and extrapolated numerical truncation. These uncertainties are quoted along with the final results while comparing with the literature values. 

\section{Results and Discussion}

We discuss first the I.P.s of the low-lying states of the Li-like systems investigated in this work. Since Ar$^{15+}$ is a highly charged ion, where relativistic and QED effects can be significant, we present results from the DC, DCB and DCQ Hamiltonians separately for Li, Be$^+$ and Ar$^{15+}$. This can demonstrate the increase in relativistic effects (especially the higher-order effects) from neutral to highly-charged systems.

In Table \ref{tabLi}, we present results for the Li atom using the above approximations in the atomic Hamiltonian from the DHF, RCCSD and full RCC methods. In addition, we also give the values from the second-order relativistic many-body perturbation theory (RMP(2) method). As can be seen from the above table, the calculated values of the electron affinities gradually increase from their DHF values, through lower-order methods, and up to the RCC method in all three type of approximations in the Hamiltonian. The Breit interaction contributes slightly higher than the QED effects, but their magnitudes are very small. The differences between the RCCSD and RCC results are on the order of $10^{-4}$, demonstrating that high-order correlations for Li energies are more pronounced than the Breit or QED effects, which enter at the level of $< 10^{-5}$.

We present the calculated values of the electron affinity of Be$^+$ from different Hamiltonians using the aforementioned methods in Table \ref{tabBe}. The trends for Be$^+$ are found to be similar to Li, with the magnitude of the triple-excitation contribution diminished to the level of $10^{-5}$. Both the Breit and QED contributions are found to be slightly higher in magnitudes compared with the Li atom, as expected. Trends in the results from both the tables presenting for the Li atom and the Be$^+$ ion also show that the Breit and QED effects are large in the $S$ states followed by the $P_{1/2}$ states, while they are negligibly small in the $D$ states. The QED effects are almost identical in both the $P_{1/2}$ and $P_{3/2}$ states. 

In Table \ref{tabAr}, we give the energy values for Ar$^{15+}$ from the same methods that were employed to the energy calculations of Li and Be$^+$. It can be seen from this table that the trends of electron correlation contributions and the relativistic effects are very different in Ar$^{15+}$ compared to the previous discussed results of Li and Be$^+$. Though we find that there are significant differences among the results from the RMP(2) and RCCSD methods, very small differences among the results from the RCCSD and RCC methods are observed. This shows that inclusion of triple excitations is important in the neutral or singly charged Li-like systems, but they are less important in the highly charged ions. It also demonstrates that both the Breit and QED interactions are quite large in this ion, with the Breit contributions larger than the QED effects. Also, most of the QED effects are contributing through the DHF method and the electron correlations affect to their contributions only slightly. It is worth mentioning here that the QED effects are estimated in this work using an effective potential. Therefore, when higher accuracy is desired for Ar$^{15+}$, it is imperative to estimate them rigorously using a more appropriate QED method.

Adding all contributions from the DC Hamiltonian and corrections from the Breit and QED interactions (DCBQ contributions), we present the final values of the electron affinity of the first eight low-lying states of the three considered atomic systems in Table \ref{tabeng}. To compare with experimentally measured values, we have taken into account corrections from our use of infinite nuclear mass by adding the NMS and SMS contributions determined in this work, scaled with the mass of the specific isotope in question. We emphasize that it is not enough to scale the energies with the reduced mass, which accounts for the nonrelativistic NMS, as the SMS effects are quite appreciable, especially for the $nP$ levels.

For $^7$Li, we compare the ground-state ionization energy with the experimental value from Ref. \cite{2007-LiE}. For the other states, we  compare the results after combining with transition frequencies from Ref. \cite{1995-LiE}. For Be$^+$, we rely on a recent critical compilation \cite{Kramida_2005}, and refer the readers to a discussion on the determination of the ionization energy {\it therein}. For Ar$^{15+}$, an accurate experimental ionization potential is not available and so we take the ground state energy from a high accuracy calculation \cite{Sapirstein}, and measured transitions from the ground state, reviewed in \cite{2010-Saloman,2019-Paul}. In Table \ref{tabeng1}, we also compare the excitation energies of the D1 and D2 lines of Ar$^{15+}$ from the high-precision calculations using the few-body methods \cite{Sapirstein,Kozhedub} with our results, which shows excellent agreement among these results. These comparisons imply that the uncertainties arising from the numerical computations and incomplete basis functions used in the calculations are quite small. Thus, we believe that the calculated wave functions of the RCC method can be used reliably for the accurate estimation of the IS constants. 

After analyzing the energies, we discuss the calculated values of the IS constants systematically. For better understanding of the trend of these calculations from the FF, EVE and AR approaches in the RCC theory framework, we present results for the FS constants first, followed by the NMS constants, then give the SMS constants in the end. As mentioned in the previous paragraph, the calculated unperturbed wave functions in our RCC method are presumed to be very accurate as the electron affinity energies obtained using the calculated wave functions agreed quite well with the literature values. Since the unperturbed wave functions are common to the FF, EVE and AR procedures to estimate the IS constants, any differences observed among the results obtained by adopting these three approaches can indicate the shortcomings of their formulations. Therefore, the present study can serve to test the potentials of the employed approaches.

In Table \ref{FScomp}, we present the FS constants for all the investigated states of Li, Be$^+$ and Ar$^{15+}$ for the comparison. We only consider the DC Hamiltonian and evaluate the FS constants in the FF, EVE and AR approaches. Further, we give results in the DHF, RCCSD and full RCC calculations in order to show the propagation of electron correlation effects from the mean-field calculation to the more exact method through the respective approach. As can be seen from the above table, the DHF values of the FS constants from the FF approach and from the other two approaches are very different. This demonstrates that orbital relaxation effects are quite significant in the determination of the FS constants. In the FF method, which is a perturbative approach, orbital relaxation arises through the all-order core-polarization effects that are usually incorporated by the random-phase approximation (RPA). It is a well known fact that the RPA effects are implicitly present in the RCC theory when it is formulated using the DHF wave function \cite{Lind,Marten,Lind1}. Therefore, the orbital relaxation effects are embedded in the EVE and AR approaches post the DHF calculations.
We also notice that the RCCSD results are comparable for all the states of Be$^+$ and Ar$^{15+}$, whereas the results of the Li atom from the FF approach show large deviation. Whereas the RCCSD and full RCC results from the EVE and AR approaches are very close to each other, the results from these methods have significant differences with respect to the FF approach. We had seen earlier that the accuracy of energies was improved from the RCCSD method to the RCC method. Thus, the above trend clearly suggests that the FF values are highly unstable for the numerical differentiation with a particular choice of perturbative parameter $\lambda$ for the determination of the FS constants. This could be due to the fact that calculations of the FS constants are more sensitive to the radial behaviors of the calculated wave functions in the nuclear region. Perhaps this can be avoided by choosing different $\lambda$ parameter for both the RCCSD and RCC calculations, but this turns out to be a shortcoming of the FF approach for this estimation of the FS constants. We find that the results from the RCCSD method and the full RCC calculations in the EVE and AR approaches are very close to each other. This indicates convergences in the results in both the approaches, and it suggests that the contributions from the terms that are neglected from Eq. (\ref{evexp}) could be very small. This may be because of the fact that the investigated systems have only a few electrons. However, we had observed large differences between the RCCSD results from both the approaches in our earlier calculation for the indium atom \cite{Sahoo_2020}.

\begin{table*}[t]
\caption{Comparison of total MS constants in GHz amu determined in this work, with other theories. The comparison is made on the same order in $\alpha$. The estimated uncertainties are given in parenthesis when they are not negligible compared with higher orders. Anomalies in the results compared with other calculations is highlighted in bold font and discussed in the text.}
\begin{ruledtabular}
\begin{tabular}{l rrcr c}
Transition & EVE-RCC & AR-RCC &  SMS(AR)+NMS(Scaling) &  Others & Ref.\\
\hline \\
& & \\
\multicolumn{6}{c}{\underline{$^{6,7}$Li}} \\
D1 (DCB)      & 443(2) & 443(2) &  443(1) & 443.986 & \cite{2017-Drake}\\
\quad QED Correction     &  &  -0.2~     &         &  0.002 &\cite{2017-Drake}  \\\\

D2 (DCB)  & \textbf{448(2)} & 443(2) &  443(1) & 444.002 & \cite{2017-Drake}\\
\quad QED Correction     &  &  -0.2~     &         &  0.002 &\cite{2017-Drake} \\\\

2S-3S (DCB)   &479(2)& 480(2) &  482.0(5) & 482.880 &  \cite{2017-Drake}\\
\quad QED Correction     &  &  -0.1~     &         &  -0.002 & \cite{2017-Drake}\\\\

2S I.P (DCB)  &757(2)&  758(2) &  760.6(5) & 761.654 &  \cite{2008-Pach}\\
\quad QED Correction     &  &  -0.2~     &         &  -0.003 & \\\\


2S-3D (DCB)    &556(2) & 558(2)    &  560.0(5)  &  561.1(5)  & \cite{1995-Yan}\\
\quad QED Correction     &  &  -0.2~     &         &   & \\\\

\multicolumn{6}{c}{\underline{$^{9,10}$Be$^+$}} \\
D1 (DCB)      & 1558(5) & 1558(5) &  1557(3) & 1559.43 & \cite{2008-Pach}\\
\quad QED Correction     &  &  -2     &         &  0.04 & \cite{2008-Pach}\\\\

D2 (DCB)      & \textbf{1577(5)} & 1558(5) &  1558(3) & 1559.62 & \cite{2008-Yan,2008-Pach}\\
\quad QED Correction     &  &  -1     &         &  0.04 & \cite{2008-Pach} \\\\
2S-3S (DCB)    & 1530(5) & 1532(5) &  1536(1) & 1537.139  & \cite{2008-Pach}\\
\quad QED Correction     &  &  -1     &         &  0.004 & \cite{2008-Pach}\\\\ 

2S I.P (DCB)  & 2524(4) & 2527(4) &  2531(1) & 2532.52 & \cite{2008-Pach} \\
\quad QED Correction     &  &  -1     &         &  0.04 & \cite{2008-Pach}\\\\
\\
\multicolumn{6}{c}{\underline{Ar$^{15^+} \times 10^3$}} \\
D1 (DCB)    & 77.7(4) & 77.7(4) & 77.6(2) & 76.69  ~ ~  & \cite{2014-Sto}\\
\quad QED Correction & &   \textbf{-0.0} &       &  0.47(5)    & \cite{2014-Sto}\\\\
D2 (DCB)    & 77.8(4) & 77.8(4) & 77.8(2) & 77.26  ~ ~  & \cite{2014-Sto}\\
\quad QED Correction & &   \textbf{-0.0} &       &   0.46(5)    & \cite{2014-Sto}\\\\  
D2-D1  (DCB)     & &    0.0(6) & ~~  0.2(3) & 0.56,0.561  & \cite{2014-Sto}, \cite{2020-PahFS}\\ 
\end{tabular}
\end{ruledtabular}
\label{MSanl}
\end{table*} 

From the above analysis, we believe that the RCC results obtained by the AR approach are more reliable owing to the fact that it evaluates the property expression more accurately. To determine the final results for the FS constants, we also performed calculations using the DCB and DCQ Hamiltonians in the AR approach. To show the change in the results due to these higher-order relativistic effects, we present results at the DHF, RCCSD and RCC level of calculations using the above Hamiltonians. These results are quoted in Table \ref{FSfin}. After accounting for the corrections from the DCB and DCQ Hamiltonians to the contributions from the DC Hamiltonian, the DCBQ values are listed in the above table. The corrections from the Breit interaction to the FS constants are larger than those from the QED effects, and amount to 1\% for Ar$^{15+}$.

We now discuss the results of the NMS constants, given in Table \ref{NMScomp}. As can be seen from the table, and unlike in the case of FS constants discussed earlier, the NMS constants for Li and Ar$^{15+}$ from the FF approach agree reasonably well with the values obtained using the EVE and AR approaches. However, there are drastic changes in some of the FF values of Be$^+$ from the RCCSD method to the full RCC calculations, which was unexpected. It may be possible to minimize such differences by changing the perturbative parameter considered in our calculation, but this ascertains our earlier finding that the results from the FF approach become highly dependent on the choice of the $\lambda$ value. The differences of the results among the EVE and AR approaches are in general very small, excluding the $2P_{1/2}$ level, where the EVE value deviates slightly from the FF and AR values, which agree with each other.

On the reasons mentioned earlier, we consider the RCC results from the AR approach as the most reliable. The DCBQ values of the NMS constants of the first eight low-lying states of the aforementioned systems are listed in Table \ref{NMSfin}. In the case of NMS constants, we find that both the Breit and QED interactions contribute less to Li and Be$^+$, whereas they are small but not insignificant in the Ar$^+$ ion. This also asserts that QED effects contribute to the NMS constants of the $S$ states more than to the other states. It could be because of the fact that the NMS constants are related to the kinetic energy of the electrons in an atomic system.
In Table \ref{NMSfin}, we also give the values obtained from the scaling law. In all levels of the considered three systems, agreement is found between scaling and our calculation to below $<0.5\%$. This fact may not be surprising for Li and Be$^+$, where relativistic effects play a smaller role, but may be considered surprising for a highly charged ion like Ar$^{15+}$. The resilience of the scaling law for the D1/D2 transitions in these systems is also demonstrated in \cite{2010-Sohlker,Li}.

Finally, we discuss the SMS constants of the undertaken Li-like systems, given in table \ref{SMScomp}. These are generally considered to be more challenging to determine accurately than the other two constants owing to the two-body form of operator. As expected, the $nP$-states exhibit larger correlation effects, while orbital relaxation effects are found to be of moderate size. Here, results for all the three approaches show a converging trend, and the discrepancies are minute. For Be$^+$ and Ar$^{15+}$, the results for the $nS$ states in the FF approach differ; otherwise there is a good agreement in the results among all other states. In fact, there is better agreement in the results between the FF and AR approaches, than with the EVE approach, which shows small but significant deviations for the $nP$-states. Adding the Breit and QED contributions to the results from the AR approach, we quote the DCBQ values for the SMS constants of the considered states of Li, Be$^+$ and Ar$^{15+}$ in Table \ref{SMSfin}. In the same table, we also give results from the DCB and DCQ Hamiltonians. From the comparison of the results from the DCB Hamiltonian and DCQ Hamiltonian with the DC Hamiltonian, we find that the Breit interaction contribute to some extent to the evaluation of the SMS constants of the ground level and contributions from the QED effects are comparably very small.

Having assessed the stability and differences in the results from the three many-body approaches of the RCC theory, we now turn to compare our results with those of very accurate theories, which are available for such systems. Using the FS constants of the individual states listed in the aforementioned tables, we give their differential values in Table \ref{FSanl} for the transitions in which very accurate calculations exist.
In general the values reported in this work for both the EVE and AR approaches are very close $<1\%$ to those of more accurate theories. We note that we evaluate the FS constants up to the first-order in $\delta \langle r^2 \rangle$, while some earlier calculations also include contributions from the higher moments. Therefore, we have distinguished the first-order contributions from $\delta \langle r^2 \rangle$ to the FS constants for the comparison purpose. This emphasizes that for such systems, a determination of the FS constants to $<1\%$ must include a discussion of such corrections. For Ar$^{15+}$, we compare our results with \cite{2014-Sto}. These are found to agree extremely well on the Coulomb and Breit interactions (here together quoted as Breit contribution), and deviate by $0.7\%$ when the QED corrections are included. As both of this work and \cite{2014-Sto} used crude method to estimate these corrections, we emphasize that if higher accuracy is required, more refined QED corrections should be investigated.

In Table \ref{MSanl}, we compare the results of our MS calculations with the values reported in high-precision calculations using few-body methods, which quote the total recoil and not the individual (SMS or NMS) contributions.
We consider the total MS obtained within the EVE and AR methods, as well as the procedure adopted in most of the IS literature, which is to combine the calculated SMS with the NMS given by the scaling law. For Li and Be$^+$, we find the combined value of SMS from AR-RCC and the NMS values from scaling to be the closest to the accurate values, followed by the calculated values from AR. The EVE MS is in most cases close and slightly less accurate than AR. For the D1/D2 transitions, the EVE SMS deviates from its accurate value. An effect which is compensated for the D1 transitions by an opposite deviation of the EVE NMS values. For Ar$^{15+}$, an agreement between the different methods in the Breit level is found to the level of one percent.

We observe large differences between the QED corrections estimated by us with the calculations carried out using the few-body methods. Whereas for Li and Be$^+$ these differences are negligible in respect to our numerical accuracy and neglected second-order mass shifts, they are quite large in Ar$^{15+}$.

\section{Conclusion}

We employed the relativistic coupled-cluster theory in three different procedures; namely finite-field, expectation value evaluation, and analytical response approaches, to determine the isotope shift constants of low-lying states of the lithium atom, and lithium-like beryllium and argon ions. Results are also given by approximating atomic Hamiltonian to the Dirac-Coulomb, Dirac-Coulomb-Breit and Dirac-Coulomb-QED Hamiltonians separately using the analytical response approach. The differences among the results from these approximated Hamiltonians demonstrate the importance of the relativistic effects to the isotope shift constants in the above atomic systems. The trend of the electron correlation effects were investigated by analysing results with respect to the mean-field calculations, and considering singles and doubles approximations in the relativistic coupled-cluster theory, and from the exact calculations. We found that the electron correlation trends are completely different in the considered three systems. Among the aforementioned three approaches, results from the finite-field approach are observed to be the least reliable. Though the starting point of the expectation value evaluation and analytical response approaches are the same, we notice significant differences in the isotope shift constants even in the full relativistic coupled-cluster calculations. By comparing with accurate results from few-body methods, we surmise that the analytical response method is more reliable and accurate than the expectation value evaluation approach for mass-shift constants, and comparable for the field shifts. As these many-body methods are suited for systems with more electrons, where few-body methods cannot be applied, our analysis suggests that the analytical response approach in the relativistic coupled-cluster theory framework can be employed to determine the isotope shift constants more reliably than the other two approaches. Lastly, we found that when QED corrections for the isotope shifts constants are desired, such as for precise calculations in highly charged ions, it does not suffice to evaluate at the Breit-level isotope shift operators. So it may require higher-order corrections to be accounted for.

\section*{Acknowledgement}

B.K.S. acknowledges use of the Vikram-100 HPC cluster of Physical Research Laboratory, Ahmedabad for carrying out the computations.

\end{document}